\documentclass[twocolumn,prl,showpacs]{revtex4-1}
\usepackage[latin9]{inputenc}
\usepackage{amsmath}
\usepackage{amssymb}
\usepackage{graphicx}
\usepackage{wasysym}

\makeatletter
\usepackage{bm}

\usepackage[]{sidecap}
\sidecaptionvpos{figure}{c}

\makeatother

\begin{document}

\title{Transmission across a bilayer graphene region}

\author{Hadi Z. Olyaei$^{1}$, Pedro Ribeiro$^{1,2}$, Eduardo V. Castro$^{1,2,3}$}

\affiliation{$^{1}$CeFEMA, Instituto Superior Técnico, Universidade de Lisboa,
Av. Rovisco Pais, 1049-001 Lisboa, Portugal}

\affiliation{$^{2}$Beijing Computational Science Research Center, Beijing 100084,
China}

\affiliation{$^{3}$Centro de F\'{\i}sica das Universidades do Minho e Porto,
Departamento de F\'{\i}sica e Astronomia, Faculdade de Ciências, Universidade
do Porto, 4169-007 Porto, Portugal}
\begin{abstract}
The transmission across a graphene bilayer region is calculated for
two different types of connections to monolayer leads. A transfer
matrix algorithm based on a tight binding model is developed to obtain
the ballistic transmission beyond linear response. The two configurations
are found to behave similarly when no gate voltage is applied. For
a finite gate voltage, both develop a conductance gap characteristic
of a biased bilayer, but only one shows a pronounced conductance step
at the gap edge. A gate voltage domain wall applied to the bilayer
region renders the conductance of the two configurations similar.
For a microstructure consisting of equally spaced domain walls, we
find a high sensitivity to the domain size. This is attributed to
the presence of topologically protected in-gap states localized at
domain walls, which hybridize as the domain size becomes of the order
of their confining scale. Our results show that transmission through
a bilayer region can be manipulated by a gate voltage in ways not
previously anticipated.
\end{abstract}
\maketitle

\section{Introduction}

The unique band structure of graphene gives rise to several alluring
phenomena which have been the subject of intense research since its
experimental discovery in 2004 \citep{Novoselov2004,Geim2007,Geim2009,CastroNeto2009}.
In particular, its high charge-carrier mobility has rendered graphene
a highly attractive and promising component for electronic and optoelectronic
devices \citep{Avouris2010,Xia2013}. Another appealing feature of
graphene for device application is its stability at the nanometer
scale, ensured by the covalent bonds among the carbon atoms \citep{Chen2008},
which is highly desirable for device-miniaturization. A graphene-based
electronic device, entirely made out of micro-structured graphene
sheets, is thus expected to reduce significantly energy dissipation
and optimize device-miniaturization and functionality \citep{Eda2009a,Georgiou2012,Jang2016}.
The recent realization of a short channel field-effect transistor,
using just 9- and 13-atom wide graphene nanoribbons \citep{Llinas2017},
is a convincing step in that direction. This is to be contrasted with
mainstream semiconductor technology which usually integrates different
materials and where component-interfacing can be difficult to scale-down
\citep{Schwierz2010}. 

Although a gapless conductor, the versatility of the electronic properties
of graphene make it possible to easily induce a gap. This can be done
by several means: cutting it into nanoribbons with zigzag or armchair
edges \citep{Son2006,Han2007,CLR+07,Wakabayashi2009}; by breaking
inversion symmetry with an appropriate substrate \citep{ZGF+07};
or applying an out-of-plain electric filed in graphene bilayer structures
\citep{Castro2007,Castro2008,Castro2010,Zhang2009}. 

Compared to monolayer graphene, the possibility of tuning the induced
gap by an external, perpendicular electrical field, which is easily
introduced through a gate potential, makes the bilayer more suitable
for device applications \citep{MK13}. Not only the gap can be tuned
by a gate bias, but also a twist angle can be engineered between the
two layers \citep{LLLdS+09,LLR+10}. This leads to a strong reconstruction
of the band structure at low energies \citep{Amorim2018}. The recent
observation of superconductivity and insulating behavior in twisted
bilayer graphene at the magic angles clearly shows the high degree
of tunability of this system \citep{Cao2018,Cao2018a}. Further manipulation
of the bilayer response is possible by inserting an insulator between
the two graphene layers, out of which tunnel field effect transistors
have been realized \citep{geimFalko14,Georgiou2012,BGJ11,britnell2012electron,Amorim2016,Chen2017a}. 

One other advantage of the graphene bilayer is that its electronic
structure of can manipulated by a layer-selective potential, induced
by a gate voltage. The possibility of sharply reversing the sign of
voltage, thus creating a well-defined one-dimensional boundary separating
regions of constant potential, has been demonstrated recently \citep{Li2016}.
These domain walls support confined one-dimensional states that are
topologically protected and can be used as purely one-dimensional
channels \citep{Martin2008,Koshino2013}. 

The ballistic transport across a bilayer graphene region has been
studied at length \citep{Snyman2007,Nilsson2007,Barbier2009,Nakanishi2010,Gonzalez2010,Gonzalez2011,Chu2017}.
Particular attention has been given to a setup where a gate voltage
is applied within the bilayer region \citep{Nilsson2007,Barbier2009,Gonzalez2011,Chu2017}.
These studies already revealed a high degree of tunability of the
transport properties. However, the effects of further manipulations
of the gate voltage, namely through the creation of a domain wall
affecting the bilayer region \citep{Li2016} are yet to be investigated.
Furthermore, the possibility of a microstructured gate voltage with
several built-in domain walls opens up new avenues to engineer electronic
transport at the nanoscale. 

The aim of this paper is to study ballistic transport of micro-structured
bilayer graphene flakes of with different types of connection to monolayer.
Using a tight-binding model of an AB staked bilayer flake, taken to
be infinite in the transverse direction, we observe that the conductance
displays aperiodic oscillations as a function of chemical potential.
The conductance in the presence of a single voltage domain is shown
to be compatible with previous results obtained within a low energy
approximation. We compute the conductance in the presence of a domain
wall in the gate bias and show that, in this case, geometries with
different types of connection to monolayer leads behave similarly.
We further study the effect of a micro-structured gate bias with multiple
domain walls. By changing the separation between domain-walls we explore
the crossover form well separated domain-wall states to the fully
hybridized regime where in-gap states start to contribute to the conductance.
Finally, we have studied the viability of an integrated nano-transistor
for experimentally reasonable conditions finding that this setup can
achieve on/off ratios of the output current within $50\lesssim I_{\text{on}}/I_{\text{off}}\lesssim200$. 

The structure of the paper is as follows: in Sec.~\ref{sec:Model-and-methods},
we introduce the model of the physical setup and the corresponding
tight-binding formulation as well as the method for obtaining the
transmission across the bilayer region using the transfer matrix formulation.
In Sec.~\ref{sec:Results-and-Discussion}, some representative results
of transmission are presented, including the new types of micro-structured
gated bilayer graphene. Sec.~\ref{sec:Conclusion} contains a short
summary and the conclusions. In Sec. \ref{secap:AppendixA}, we present
some of the details of the calculation of the transmission. 

\section{Model and methods\label{sec:Model-and-methods}}

\begin{figure}
\begin{centering}
\includegraphics[width=1\columnwidth]{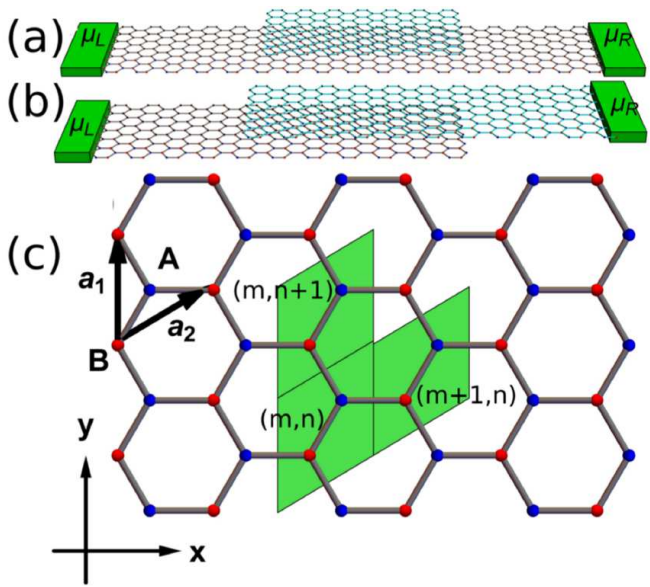}
\par\end{centering}
\centering{}\caption{\label{model}System setups with a bilayer region used in this work:
(a) the $1\rightarrow1$ setup; (b) the $1\rightarrow2$ setup. (c)~Primitive
vectors $\boldsymbol{a}_{1}$ and $\boldsymbol{a}_{2}$, sublattice
labels $A$ (blue) and $B$ (red), and unit cell labeling of the monolayer
strucuture.}
\end{figure}

Schematics of the setup for which the transmission and the conductance
are studied is shown in Fig.~\eqref{model}. The case of Fig.~\ref{model}(a)
consists of a single layer graphene with a flake of another layer
on top, the $1\rightarrow1$ setup. The second configuration is obtained
from two sheets of graphene that are partially overlapped, the $1\rightarrow2$
setup, as shown in Fig.~\ref{model}(b). In both we consider A-B
stacking. Translational invariance along the transverse direction
($y$-axis) is presumed. We are interested in the ballistic regime
where the electronic mean free path is larger than the typical length
of the device. For simplicity, we consider the case of perfect contacts,
which can be replaced by infinite leads.

We model electrons in the structure using the conventional tight-binding
approach for $p_{z}-$electrons \citep{CastroNeto2009} hopping between
nearest neighbor carbon sites of the atomic lattice shown in Fig.~\eqref{model}(c),
which can be written as $H=H_{1}+H_{2}+H_{\perp}$. Here, 

\begin{align}
H_{j} & =-t\sum_{m,n}a_{j,m,n}^{\dagger}\left[b_{j,m,n}+b_{j,m+1,n}+b_{j,m,n+1}\right]\label{eq:hamiltonian}\\
 & +\sum_{m,n\in\text{BL}}V_{j}\left[a_{j,m,n}^{\dagger}a_{j,m,n}+b_{j,m,n}^{\dagger}b_{j,m,n}\right]+h.c
\end{align}
is the Hamiltonian of the $j=1,2$ layer, and 
\begin{equation}
H_{\perp}=-t_{\perp}\sum_{m,n\in\text{BL}}a_{1,m,n}^{\dagger}b_{2,m,n}+h.c\,.
\end{equation}
is the inter-layer hopping term, with $a_{j}^{\dagger}\left(b_{j}^{\dagger}\right)$
the creation operators of a particle in sublattice $A\left(B\right)$
in the $\left(m,n\right)$ unit cell of the $j$th layer. The effect
of an applied gate voltage within the bilayer region is modeled by
$V_{j}$. The $\text{BL}$ restriction in the summation stands for
sites belonging to the bilayer region. 

\begin{figure}
\begin{centering}
\includegraphics[width=1\columnwidth]{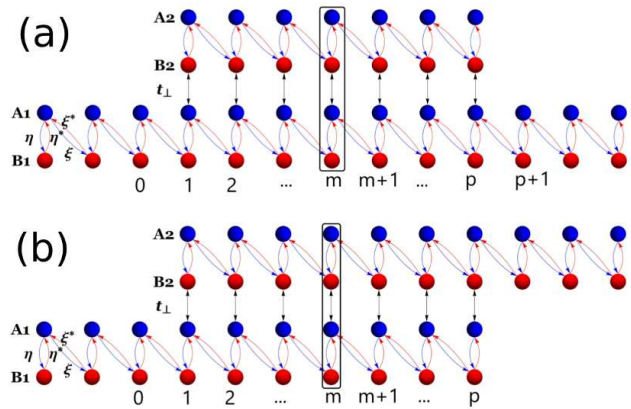}
\par\end{centering}
\caption{\label{fig:1D-effective-chain}1D effective chain obtained after Fourier
transform, as described in the main text: (a) The $1\rightarrow1$
configuration; (b) the $1\rightarrow2$ configuration.}
\end{figure}

After Fourier transformation in the $y-$direction, the stationary
states of the 1D effective chain for the two cases shown in Fig.~\eqref{fig:1D-effective-chain}
can be written as 

\begin{equation}
|\psi_{k}\rangle=\sum_{j=1,2}\sum_{m}\left(\psi_{m,k}^{A,j}a_{j,m,k}^{\dagger}+\psi_{m,k}^{B,j}b_{j,m,k}^{\dagger}\right)|0\rangle\,,\label{eq:general solution}
\end{equation}
with $k$ the wave number along the $y$ direction. Within the monolayer
(lead) region, we define the column vector $\Psi_{k}^{j}\left(m\right)=\left[\begin{array}{cc}
\psi_{m-1,k}^{A,j} & ,\psi_{m,k}^{B,j}\end{array}\right]^{T}$ which obeys the transfer matrix equation (see Appendix~\ref{secap:AppendixA}),

\begin{equation}
\Psi_{k}^{1\left(2\right)}\left(m+1\right)=\mathbb{T}_{L}\Psi_{k}^{1\left(2\right)}\left(m\right)\,,\label{tmatrix for leads}
\end{equation}
where $\mathbb{T}_{L}$ is given by

\begin{equation}
\mathbb{T}_{L}=\frac{1}{\xi\eta^{\ast}}\left[\begin{array}{cc}
-|\xi|^{2} & -\epsilon\xi\\
\epsilon\xi^{\ast} & \epsilon^{2}-|\eta|^{2}
\end{array}\right]\,,
\end{equation}
with $\eta=t\left(1+e^{ik}\right)$ and $\xi_{k}=t$. For the bilayer
region ($1\leq m\leq p$) we define $\boldsymbol{\Psi}_{k}\left(m\right)=\left[\begin{array}{cc}
\Psi_{k}^{1}\left(m\right) & \Psi_{k}^{2}\left(m\right)\end{array}\right]^{T}$ obeying 

\begin{equation}
\boldsymbol{\Psi}_{k}\left(m+1\right)=\mathbb{T}_{BL}\boldsymbol{\Psi}_{k}\left(m\right)\,,\label{eq:tmatrix for bilayer}
\end{equation}
where the transfer matrix $\mathbb{T}_{BL}$ is given by

\begin{align}
\mathbb{T}_{BL} & =\frac{1}{\eta^{\ast}\xi}\left[\begin{array}{cccc}
-|\eta|^{2} & -\epsilon\eta & 0 & 0\\
\epsilon\eta^{\ast} & \epsilon^{2}-|\xi|^{2} & 0 & -t_{\perp}\\
t_{\perp}\frac{|\eta|^{2}}{\xi^{\ast}} & t_{\perp}\frac{\epsilon\eta}{\xi^{\ast}} & -|\eta|^{2} & -\epsilon\eta\\
-t_{\perp}\frac{\epsilon\eta^{\ast}}{\xi^{\ast}} & -t_{\perp}\frac{\epsilon^{2}}{\xi^{\ast}} & \epsilon\eta^{\ast} & \epsilon^{2}-|\xi|^{2}
\end{array}\right]\,.
\end{align}
The amplitudes at the left and the right interfaces can be related
by,

\begin{equation}
\boldsymbol{\Psi}_{k}\left(p+1\right)=\left(\mathbb{T}_{BL}\right)^{p}\boldsymbol{\Psi}_{k}\left(1\right)\label{M matrix}
\end{equation}
and by the boundary conditions: $\psi_{0,k}^{A2}=\psi_{p+1,k}^{B2}=0$
for the $1\rightarrow1$ case, and $\psi_{0,k}^{A2}=\psi_{p+1,k}^{B1}=0$
for the $1\rightarrow2$ case, as can be seen in Figs.~\ref{fig:1D-effective-chain}(a)
and \ref{fig:1D-effective-chain}(b). With these boundary conditions
one obtains the matrix $\mathbb{\mathcal{M}}_{1\rightarrow1\left(2\right)}$
relating the layer 1 in the left to layer 1(2) in the right,

\begin{equation}
\Psi_{k}^{1\left(2\right)}\left(p+1\right)=\mathbb{\mathcal{M}}_{1\rightarrow1\left(2\right)}\Psi_{k}^{1}\left(1\right)\,,\label{eq:left edge to right edge}
\end{equation}
where $\mathbb{\mathcal{M}}_{1\rightarrow1\left(2\right)}$ are defined
from Eq.~\eqref{M matrix} in Appendix~\ref{secap:AppendixA}.

Within the semi-infinite leads, Eq.~\eqref{tmatrix for leads} can
be solved by assuming the ansatz
\[
\Psi_{k}^{1\left(2\right)}\left(m\right)=\alpha_{+}\lambda_{+}^{m-1}\zeta_{k}^{+}+\alpha_{-}\lambda_{-}^{m-1}\zeta_{k}^{-}\,,
\]
with $\mathbb{T}_{L}\zeta_{k}^{\pm}=\lambda_{\pm}\zeta_{k}^{\pm}$.
The eigenvalues $\lambda_{\pm}$ and the eigenmodes $\zeta_{k}^{\pm}$
are explicitly derived in Appendix~\ref{secap:AppendixA}. In the
leads we only consider propagating modes, so that $|\lambda|=1$.
The eigenmodes are thus interpreted as left-moving, $\zeta_{j}^{-}$,
and right-moving, $\zeta_{j}^{+}$, modes, according to their group
velocity (see Appendix~\ref{secap:AppendixA}). We then use the $\zeta_{k}^{\pm}$
eigenbasis to write the wave function in the leads, 
\begin{equation}
\Psi_{k}^{1\left(2\right)}\left(m\right)=\lambda_{+}^{m-1}\zeta_{k}^{+}+\lambda_{-}^{m-1}r_{k}\zeta_{k}^{-}\quad,\quad m<1\label{eq:leftlead}
\end{equation}

\begin{equation}
\Psi_{k}^{1\left(2\right)}\left(m\right)=\lambda_{+}^{m-p-1}\tau_{k}\zeta_{k}^{+}\quad,\quad m>p\,,\label{eq: rightlead}
\end{equation}
from which we define transmission and reflection coefficients, respectively
$\tau$ and $r$. The transmission and reflection coefficients are
given by, 

\begin{equation}
\left[\begin{array}{c}
\tau_{k}\\
0
\end{array}\right]=U^{-1}\mathcal{M}_{1\rightarrow1\left(2\right)}U\left[\begin{array}{c}
1\\
r_{k}
\end{array}\right]\,,
\end{equation}
where $U=\left[\begin{array}{cc}
\zeta_{k}^{+}, & \zeta_{k}^{-}\end{array}\right]$.

The transmission probability is then defined as $T\left(\epsilon,k\right),=1-|r_{k}|^{2}=|\tau_{k}|^{2}$,
and the overall transmission per transverse unit length is given by, 

\begin{equation}
\bar{T}\left(\epsilon\right)=\frac{1}{2\pi}\int_{-\pi}^{\pi}dkT\left(\epsilon,k\right)\,.\label{Average T}
\end{equation}
Using the Landauer formula \citep{Datta2005a}, we find the current
per transverse unit length across the bilayer region,

\begin{equation}
I=\frac{2e}{h}\int d\epsilon\bar{T}\left(\epsilon\right)\left[f\left(\epsilon-\mu_{L}\right)-f\left(\epsilon-\mu_{R}\right)\right]\label{LB formula-1}
\end{equation}
where $f(\epsilon)$ is the Fermi distribution function and $\mu_{L}$($\mu_{R}$)
are the chemical potential in the left (right) lead (in the following
we assume $\mu_{L}>\mu_{R}$). Assuming $\mu\equiv\mu_{L}=\mu_{R}+\delta\mu$,
with $\delta\mu\ll\mu$, we can linearize the Landauer formula \citep{Buttiker1985}
to obtain the conductance $G\equiv e\delta I/\delta\mu$, which can
be written as

\begin{equation}
G\left(\mu\right)=-G_{0}\int d\epsilon T\left(\epsilon\right)\frac{\partial f\left(\epsilon-\mu\right)}{\partial\epsilon}\,,\label{linearLformula}
\end{equation}
where $G_{0}=\frac{2e^{2}}{h}$ is the conductance quantum. For a
system at zero temperature, Eq.~\eqref{linearLformula} can be simplified
to $G=G_{0}T\left(\mu\right)$.

\section{Results and Discussion\label{sec:Results-and-Discussion}}

\begin{figure}
\begin{centering}
\includegraphics[width=1\columnwidth]{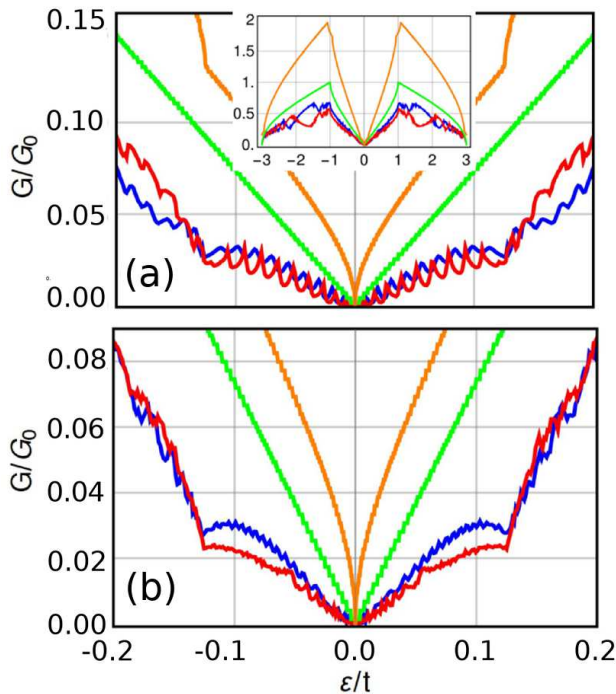}
\par\end{centering}
\centering{}\caption{\label{fig:Graphene condactance} Transmission per transverse unit
length near the Fermi-level for the $1\rightarrow1$ (blue), $1\rightarrow2$
(red) geometries plotted for $p=200$ (a) and $p=1500$ (b). The transmission
for an infinite graphene layer (green) and for an infinite bilayer
(orange) are plotted for comparison Inset: Transmission for the whole
bandwidth for $p=200$.\textbf{ }}
\end{figure}

\subsection{Transmission through a bilayer graphene region}

In this section, we compute the transmission amplitudes for the $1\rightarrow1$
and $1\rightarrow2$ cases. A simplifying feature is that, for both
cases, there is only one propagating incident mode associated with
given $\epsilon$, hence the corresponding transfer matrix of the
leads is a two by two. Note that, due to electron-hole symmetry, $T\left(\epsilon,k\right)=T\left(\pm\epsilon,k\right)$. 

Figure~\ref{fig:Graphene condactance} shows the conductance for
energies near the Fermi-level for the $1\rightarrow1$ (blue) and
$1\rightarrow2$ (red) geometries and for two values of the scattering
region size, $p=200$, (a), and for $p=1500$, (b). For comparison,
the conductance through an infinite system consisting of a single
(green) or a double (orange) graphene layer is also depicted. Note
that, in these cases the total transmission in Eq.~\eqref{Average T}
is simply determined by the dispersion relation. Therefore, for low
energies it behaves as $\propto\left|\varepsilon\right|$ for the
single layer and as $\propto\left|\varepsilon\right|^{1/2}$ for the
bilayer. 

For both geometries, the low energy conductance is almost twice as
low as for pristine graphene and vanishes faster, with a $\propto\left|\varepsilon\right|^{2}$
scaling behavior. The inset of Fig.~\ref{fig:Graphene condactance}(a),
depicting $G$ for the all energies within the bandwidth, shows that,
even away from the Fermi-level, $G$ never attains the value of the
pristine case. 

Another pronounced low energy feature of the transmission, is the
sudden increase for energies around $t_{\perp}$. Thus, as also seen
in the pristine double layer case, the conductance resolves the appearance
of the higher energy band, after which two propagating modes become
available for transport within the bilayer region. 

The differences between the $1\rightarrow1$ and $1\rightarrow2$
geometries are more pronounced for higher energies. At low energies,
they can be completely masked out by the finite-size effects that
yield the characteristic jumps in the conductance, Fig.~\ref{fig:Graphene condactance}(a).
For larger values of $p$, when the finite-size oscillations are reduced,
the $1\rightarrow1$ case is seen to have a higher conductance. This
is to be expected since in this case, the transmitted electrons do
not have to change layer, which is suppressed for low values of $t_{\perp}$. 

\subsection{Conductance through a \emph{gated} bilayer graphene region}

\begin{figure*}
\begin{centering}
\includegraphics[width=1\textwidth]{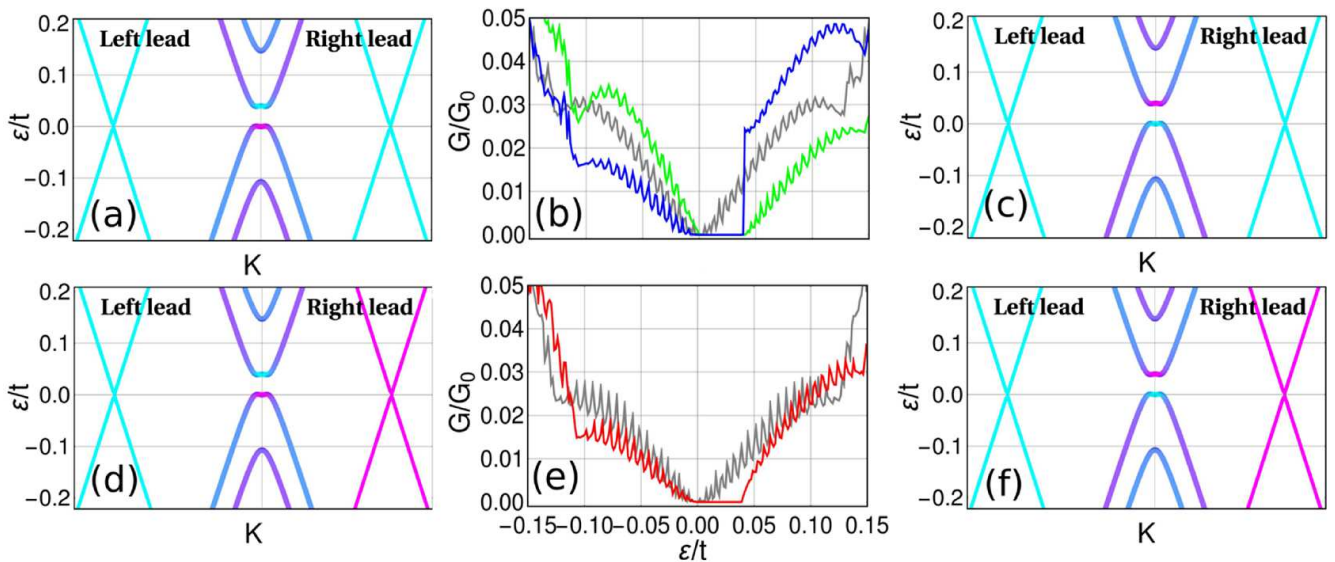}
\par\end{centering}
\centering{}\caption{\label{fig:transmission_finite_V}Conductance for a gated bilayer
region. Upper panels refer to the geometry $1\rightarrow1$ (a-c)
and lower panels to $1\rightarrow2$ (d-f). (b) Blue and (e) red correspond
to $V_{1}=0.04t$ and $V_{2}=0$, computed for $p=400$. For (b) green
and (e) red: the values of the voltage are swapped, i.e. $V_{1}=0$
and $V_{2}=0.04t$. Notice that the geometry $1\rightarrow2$ is unchanged
under swapping the gate voltage. The unbiased case $V_{1}=V_{2}=0$
is depicted as a gray line for comparison. The dispersion relations
at low energies, computed for an infinite system, corresponding respectively
to the setups (b) blue , (b) green, and (e) red are given in (a),
(c), (d) and (f). The central dispersion corresponds to the bilayer
region and the color encodes whether the wave-function is localized
in the bottom (cyan) or in the upper (pink) layers. The left and right
dispersions correspond to a single layer and follow the same color
coding. }
\end{figure*}

\begin{figure}
\begin{centering}
\includegraphics[width=1\columnwidth]{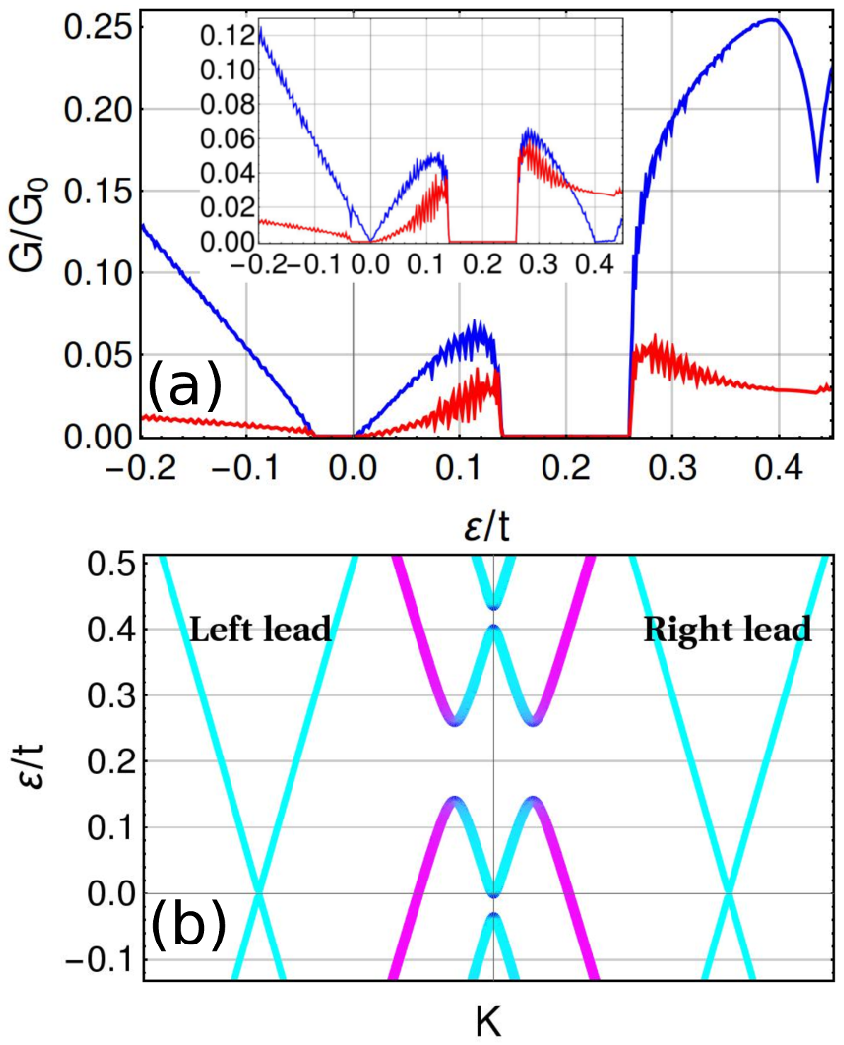}
\par\end{centering}
\caption{\label{fig:higher voltage}(a) Conductance for the a gated bilayer
region computed for $p=400$ for the $1\rightarrow1$ (blue) and $1\rightarrow2$
(red) geometries with $V_{1}=0.4t$ and $V_{2}=0$ . The inset depict
the opposite voltage configuration: $V_{1}=0$ and $V_{2}=0.4t$.
(b)~The dispersion relations at low energies, computed for an infinite
system, corresponding to the setup (a). The central dispersion corresponds
to the bilayer region and the color encodes whether the wave-function
is localized in the bottom (cyan) or in the upper (pink) layers. The
left and right dispersions correspond to a single layer and follow
the same color coding.}
\end{figure}

\subsubsection{Homogeneous case}

In this section, we study the effect on the transmission of a gate
voltage applied within the bilayer region. We assume that only one
of the layers is affected by the gate while the other remains at zero
voltage. We study the cases for which the voltage of the lower, $V_{1}$,
or upper layers, $V_{2}$, is $0.04t$ or ten times larger $0.4t$,
which correspond to typical values of gate voltages that can be implemented
experimentally.

Fig.~\ref{fig:transmission_finite_V} shows the conductance through
a gated bilayer graphene region in different cases together with a
plot of the band structure of the bilayer and the single layer leads
around zero energy (computed assuming an infinite system). 

Fig.~\ref{fig:transmission_finite_V}(b) depicts the $1\rightarrow1$
geometry for $V_{1}=0.04t$ and $V_{2}=0$ (blue) and for the swapped
voltage configuration $V_{1}=0$ and $V_{2}=0.04t$ (green). The most
pronounced features are the suppression of transport for $\varepsilon\in\left\{ 0,|\Delta V|\right\} $
and a jump in the conductance for $\varepsilon\approx|\Delta V|$
seen in~\ref{fig:transmission_finite_V}(b) blue, which is not present
when the gate voltages are swapped in~\ref{fig:transmission_finite_V}(b)
(green). The illustrations of the band structures in Figs.~\ref{fig:transmission_finite_V}(a)
and \ref{fig:transmission_finite_V}(c) help to understand this behavior.
The effect of the gate voltage is to open up a gap in the dispersion
relation of the bilayer. Moreover, while for $V_{1}=V_{2}=0$, the
wave-function's amplitudes are equally distributed between the two
layers of the bilayer system, for finite voltages their distribution
changes drastically near the gap edges (valence band maximum and conduction
band minimum). The color coding in Fig.~\ref{fig:transmission_finite_V}(a)
and~\ref{fig:transmission_finite_V}(c) shows the localization of
the wave-function in the upper or lower layers. This energy-dependent
layer distribution can simply explain the conduction jump: in the
case depicted in~\ref{fig:transmission_finite_V}(a), after passing
the energy gap the system has suddenly available a large density of
transmission modes within the lower layer. Such matching conditions
(same color, at a given energy, for the leads and the bilayer region)
never arises in the opposite case, \ref{fig:transmission_finite_V}(b)
green, as can be seen in~\ref{fig:transmission_finite_V}(c). 

Figs.~\ref{fig:transmission_finite_V}(e) depicts the transmission
for the $1\rightarrow2$ geometry. This case is symmetric under the
swapping of the voltages. In this case Figs.~\ref{fig:transmission_finite_V}(d)
and \ref{fig:transmission_finite_V}(f) show that the perfect matching
conditions seen in \ref{fig:transmission_finite_V}(a) are never attained
and thus no jump in conductance is observed. 

The conductance attained when the gate voltage is increased by one
order of magnitude is depicted in Fig.~\ref{fig:higher voltage}(a)
for the two geometries $1\to1$ (blue) and $1\to2$ (red) for $V_{1}=0.4t$
and $V_{2}=0$. The inset shows the voltage swapped case, $V_{1}=0$
and $V_{2}=0.4t$. Fig.~\ref{fig:higher voltage}(b) depicts the
band structure, with the same color coding as before, corresponding
to the case $1\to1$ and $V_{1}=0.4t$ and $V_{2}=0$. An interesting
feature of the transmission in Fig.~\ref{fig:higher voltage}(a)
is that there are two regions where the conduction seems to vanish.
One, at higher energies, corresponds to the band-gap and thus the
suppression of the conductance is not surprising. However, the second
arises within a region where the density of states is finite. Again,
the plot of the band structure in Fig.~\ref{fig:higher voltage}(b)
can simply explain this effect: the gap in conductance corresponds
to a region where the conducting states with support on the lower
layer become gapped, so although the total density of states is finite,
there are no states contributing to transport. 

\subsubsection{Inhomogeneous case: single domain wall }

\begin{figure}
\begin{centering}
\includegraphics[width=0.9\columnwidth]{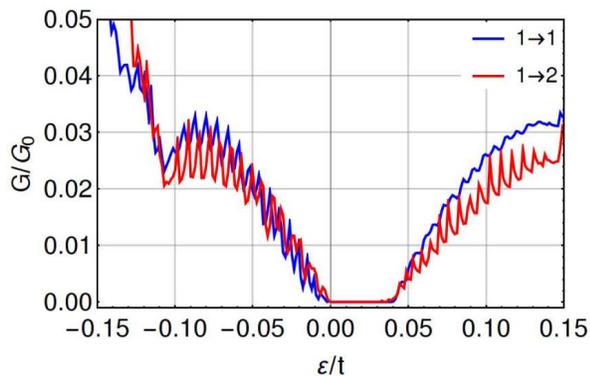}
\par\end{centering}
\caption{\label{fig:single domain wall}Conductance for the bilayer region
with a gate voltage domain wall in the middle, computed for $p=400$
in the $1\rightarrow1$ (blue) and $1\rightarrow2$ (red) geometries
with $V_{0}=0.04t$.}
\end{figure}

In this section, we study how the transmission is affected by the
presence of an inhomogeneous gate voltage. We consider the simplest
case where a gate voltage domain wall is present in the bilayer region.
We assume that the local potential at cell $m$, layer $j$ (see Fig.~\ref{fig:1D-effective-chain})
is given by $V_{m,j}=V_{0}\Theta\left[\left(-1\right)^{j}\left(m-\frac{p}{2}\right)\right]$,
with $\Theta(x)$ the Heaviside function. Therefore, the potential
difference on the left half ($m<p/2)$ is $V_{1}-V_{2}=V_{0}$ while
on the right half ($m>p/2$) it is $V_{1}-V_{2}=-V_{0}$, which implies
a domain wall right at the middle of the bilayer region. This domain
wall structure is known to support confined states, localized in the
transverse direction and extending along the wall \citep{Martin2008},
with important consequences regarding transport in the direction of
the wall \citep{Li2016}. The impact of a domain wall on charge transport
in the perpendicular direction has not been studied before and is
analyzed in the following.

In Fig.~\ref{fig:single domain wall} we show the conductance for
the geometries $1\to1$ (blue) and $1\to2$ (red) for $V_{0}=0.04t$.
The two geometries now have very similar conductance, which contrasts
with the case when no domain wall is present, depicted in Figs.~\ref{fig:transmission_finite_V}(b)
and~\ref{fig:transmission_finite_V}(f). A noticeable difference
is the absence of the jump in conductance observed for the $1\to1$
geometry in Fig.~\ref{fig:transmission_finite_V}(b). Since the domain
wall reverses the layer distribution of the wave-function's amplitudes,
the perfect matching conditions seen in \ref{fig:transmission_finite_V}(a)
are never attained and thus no jump in conductance is observed. We
conclude that the domain wall erases the difference between the two
geometries.

As shown in Ref.~\citep{Martin2008}, the states confined at the
domain wall originate one-dimensional bands dispersing inside the
bulk gap. In Fig.~\ref{fig:single domain wall} the impact of those
states is unnoticeable, as a well resolved gap of order $\sim V_{0}$
is still apparent. This can be understood as a consequence of transverse
confinement. At low energies, the wave function of these states has
a decay length of the order $\beta\approx a_{0}t/\sqrt{V_{0}t_{\perp}}\gg a_{0}$,
where $a_{0}$ is the carbon-carbon distance \citep{Martin2008}.
For $V_{0}=0.04t$ the decay length is $\beta\approx8a_{0}$, much
smaller than the distance $l=200a_{0}$ between the domain wall and
the edges of the scattering region. Therefore, for a single domain
wall, these states do not contribute in propagating charge across
the bilayer region.

\subsection{Conductance through a \emph{microstructured} biased bilayer graphene
region}

We now generalize our study to multiple domain walls. Our aim is to
show how these microstructures, that are now routinely fabricated,
can be used to engineer the transmission. We consider the potential
of the previous section generalized for a periodic gated region of
size $l$, $V_{m,1}=V_{0}\Theta_{l}\left[m\right]$, where 
\begin{align*}
\Theta_{l}\left[m\right] & =\begin{cases}
1 & \text{if }2kl/a_{0}<m<\left(2k+1\right)l/a_{0}\text{ for }k\in\mathbb{Z}\\
0 & \text{if else}
\end{cases},
\end{align*}
and $V_{m,2}=V_{0}\left(1-\Theta_{l}\left[m\right]\right)$. As a
function of $l$, there are two qualitatively different cases that
we consider in the following: a large domain length, $l\gg\beta$,
where the edge modes along the domain wall do not hybridize and thus
do not contribute to the transport properties; and a small domain
length, $l\apprle\beta$, for which there is hybridization of edge
modes and thus transport for energies within the bulk gap becomes
possible. 

\begin{figure}
\begin{centering}
\includegraphics[width=1\columnwidth]{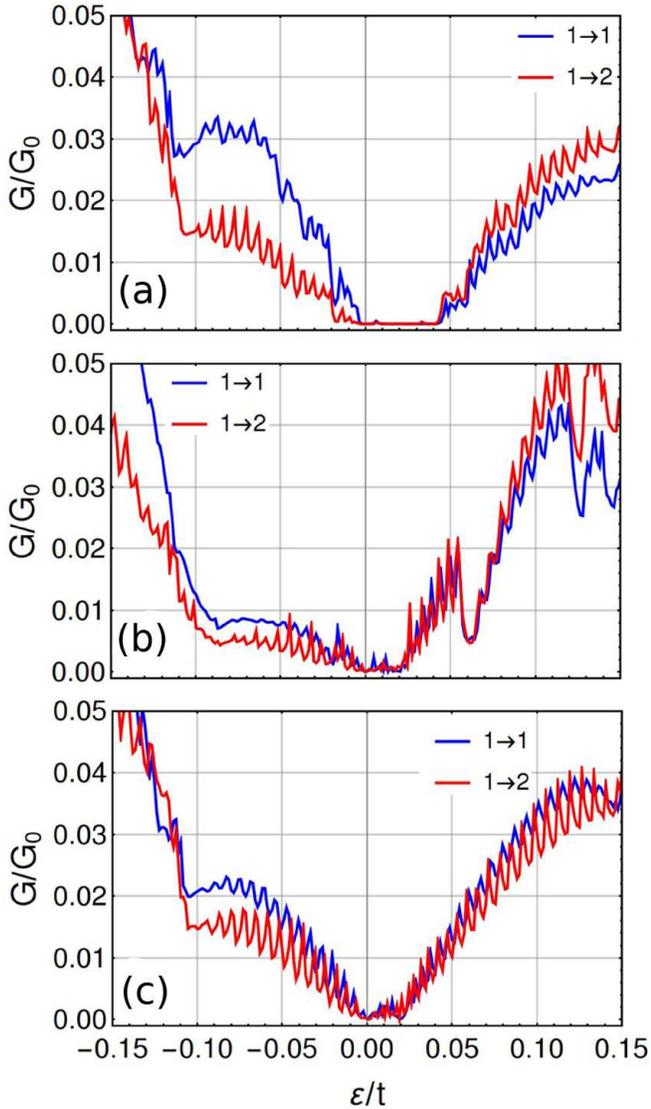}
\par\end{centering}
\caption{\label{fig:multiple domain walls}Conductance for the $1\rightarrow1$
(Blue) and the $1\rightarrow2$ (red) geometries in the presence of
multiple domain walls separated by $l$ lattice spacings, computed
for $V_{0}=0.04t$. (a) $l=80a_{0}\gg\xi$. (b) $l=20a_{0}>\xi$.
(c) $l=5a_{0}\approx\xi$.}
\end{figure}

Figure~\ref{fig:multiple domain walls} shows the evolution of the
conductance curves with $l$. We consider, as before, $V_{0}=0.04t$
corresponding to $\beta\approx8a$. In Fig.~\ref{fig:multiple domain walls}(a)
we show the conductance for $l=80a_{0}\gg\beta$. As for the $l=p/2$
case in the previous section, the differences between the two geometries
are not significant and there is almost no conductance within the
gap, for $\varepsilon\in\left\{ 0,V_{0}\right\} $. 

Figure~\ref{fig:multiple domain walls}(b) depicts the conductance
for a smaller value of $l=20a_{0}$. Here, there are already some
states within the gap that contribute to transport which result from
the hybridization of the edge modes along the domain walls. 

In Fig.~\ref{fig:multiple domain walls}(c) we set $l=5a_{0}$, for
which the domain wall states are already fully hybridized. Note the
striking similarity between the low energy conductance and that obtained
for an unbiased bilayer region, shown in Fig.~\ref{fig:Graphene condactance}
and as a background in Figs.~\ref{fig:transmission_finite_V}(b-c)
and~\ref{fig:transmission_finite_V}(f-g). It is clear that the effect
of the gap has been completely washed out. At higher energies, however,
the system still shows the conductance asymmetry typical of a gate
biased bilayer region {[}see Figs.~\ref{fig:transmission_finite_V}(b-c)
and~\ref{fig:transmission_finite_V}(f-g){]}.

\subsection{Results for current at finite temperature and device application}

In this section, we study the viability of an integrated nano-transistor
based on the $1\rightarrow1$ or $1\rightarrow2$ geometries. 

For this device, one aims to maximize the current ratio between the
``on'' and ``off'' currents, $I_{\text{on}}$ and $I_{\text{off}}$,
passing through the terminals, when changing between two values of
the applied gate voltage. Due to its low resistance and versatility,
graphene is a natural candidate for transistor implementations. However,
due to the nature of its band structure, achieving a high on/off ratio
is a technical challenge especially at finite temperature. We exploit
the non-linear behavior of the conductance obtained with the setup
of Fig.~\ref{fig:transmission_finite_V}(b) to optimize the $I_{\text{on}}/I_{\text{off}}$
and study its behavior at finite temperature. 

\begin{figure}
\begin{centering}
\includegraphics[width=1\columnwidth]{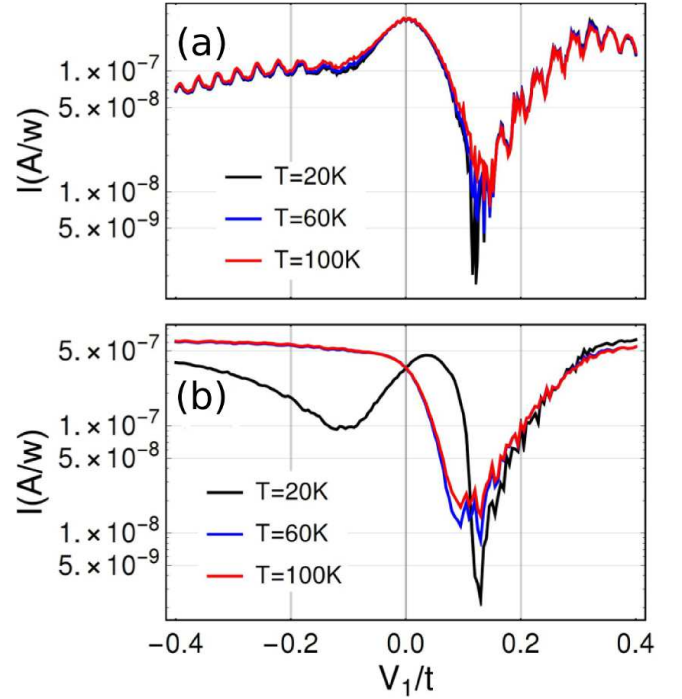}
\par\end{centering}
\caption{\label{fig:Logarithmic-density-plot}Current as a function of gate
voltage for the setup of Fig.~\ref{fig:transmission_finite_V}(b)
as a function of $V_{1}$ (for $V_{2}=0$) computed for $p=200$ and
different values of the temperature for the $1\rightarrow1$ (a) and
$1\rightarrow2$ (b) geometries.}
\end{figure}

Figure~\eqref{fig:Logarithmic-density-plot} shows the logarithmic
plot of the current for the $1\rightarrow1$, panel \eqref{fig:Logarithmic-density-plot}(a),
and $1\rightarrow2$, panel \eqref{fig:Logarithmic-density-plot}(b),
setups for different temperatures as a function of gate voltage $V_{1}$.
The chemical potentials on the left and right leads were fixed at
the experimentally reasonable values of $\mu_{\text{L}}=0.1t\text{ and }\mu_{\text{R}}=0$.
In the gate voltage interval $0<V_{1}<0.2t$ this setup can achieve
$50\lesssim I_{\text{on}}/I_{\text{off}}\lesssim200$. 

\section{Conclusion\label{sec:Conclusion}}

In this work, we have studied the conductance across a graphene bilayer
region for two different positions of the single layer leads: the
case when the leads connect to the same layer, the $1\rightarrow1$
configuration; and the case when the leads connect to different layers,
$1\rightarrow2$ configuration. We have worked in the limit of an
infinitely wide scattering region, to avoid edge effects, and developed
a transfer matrix, tight-binding based methodology which allows going
away from linear response. We have found that, when there is no gate
bias applied to the bilayer region, the two setups, $1\rightarrow1$
and $1\rightarrow2$, have a similar behavior, with a slightly higher
conductance in the $1\rightarrow1$ configuration. The presence of
a bias gate voltage differentiates between the two configurations.
Both of them develop a conductance gap which mimics the spectral gap
of a biased bilayer, but only the $1\rightarrow1$ configuration shows
a pronounced conductance step at one of the gap edges, extending the
results obtained in the continuum limit \citep{Nilsson2007} and for
ribbons of finite width \citep{Gonzalez2011}. This step is not present
if the gate polarity is reversed. Introducing a domain wall in the
gate bias applied to the bilayer region, the conductance step disappears
and the two configurations, $1\rightarrow1$ and $1\rightarrow2$,
behave again in a similar way. 

We have also studied the effect of a gate bias with a multiple domain
wall microstructure applied to the bilayer region. When the separation
between domains is much larger than the localization length of the
states confined at the domain walls, the multiple domain walls states
behave independently and the result is similar to the case of a single
domain wall. On decreasing the separation between domain walls, the
localized states start to hybridize and a finite conductance starts
to appear inside the gap. At even smaller distances, the gap is completely
washed out, and only at higher energies a conductance asymmetry characteristic
of a gate biased bilayer region is present. Finally, we have studied
the viability of an integrated nano-transistor based on the $1\rightarrow1$
or $1\rightarrow2$ geometries. For experimentally reasonable chemical
potential difference ($\sim0.3\,\text{eV}$) and gate voltage interval
(from 0 up to $\sim0.6\,\text{eV}$) we have found that this setup
can achieve $50\lesssim I_{\text{on}}/I_{\text{off}}\lesssim200$.
Summing up all the finds, it is clear the transmission through a bilayer
region can be manipulated by a gate bias in ways not previously anticipated.
\begin{acknowledgments}
We thanks B. Amorim for valuable discussions. The authors acknowledge
partial support from FCT-Portugal through Grant No. UID/CTM/04540/2013.
H.Z.O acknowledges the support from the DP-PMI and FCT (Portugal)
through scholarship PD/ BD/113649/2015. PR acknowledges support by
FCT-Portugal through the Investigador FCT contract IF/00347/2014. 
\end{acknowledgments}

\appendix

\section{The transmission through a bilayer region\label{secap:AppendixA}}

Here we detail the transfer matrix method used to obtain the transmission
coefficient. We apply Fourier transformation,
\[
a_{j,m,k}^{\dagger}\left(b_{jm,k}^{\dagger}\right)=\frac{1}{\sqrt{N_{y}}}\sum_{n}\exp\left(ikn\right)a_{j,m,n}^{\dagger}\left(b_{j,m,n}^{\dagger}\right),
\]
to the tight-binding Hamiltonian \eqref{eq:hamiltonian} and obtain
\begin{align}
H_{k} & =-\sum_{j,m}a_{j,m,k}^{\dagger}\left[\eta b_{j,m,k}+\xi b_{j,m+1,k}\right]\label{eq:hamiltonian-1}\\
 & +\sum_{j,m}^{BL}V_{j}\left[a_{j,m,k}^{\dagger}a_{j,m,k}+b_{j,m,k}^{\dagger}b_{j,m,k}\right]+\text{H.c.}\nonumber \\
 & -t_{\perp}\sum_{m}^{BL}a_{1,m,k}^{\dagger}b_{2,m,k}+\text{H.c.}\nonumber 
\end{align}
where $\eta$ and $\xi$ are defined in the main text.

By multiplying $H_{k}|\psi_{k}\rangle=\epsilon_{k}|\psi_{k}\rangle$by
$\langle m,l,\mu|$, for a given lattice point $\left(m,l,\mu\right)$,
where $m$ stands for position, $l$ for layer, and $\mu=A,B$ labels
sublattices, one obtains, for the leads where $m<0$ or $m>p+1$,

\begin{align*}
\epsilon\psi_{m,k}^{A1} & =-\eta\psi_{m,k}^{B1}-\xi\psi_{m+1,k}^{B1}\\
\epsilon\psi_{m,k}^{B1} & =-\eta^{\ast}\psi_{m,k}^{A1}-\xi^{\ast}\psi_{m-1,k}^{A1},
\end{align*}
with $\psi_{m,k}^{\mu l}=\langle m,\mu,l|\psi_{k}\rangle$. We rewrite
the latter equations in a matrix equation form as

\begin{align}
\left[\begin{array}{cc}
\epsilon & \xi\\
\eta^{\ast} & 0
\end{array}\right]\left[\begin{array}{c}
\psi_{m,k}^{A1}\\
\psi_{m+1,k}^{B1}
\end{array}\right]=- & \left[\begin{array}{cc}
0 & \eta\\
\xi^{\ast} & \epsilon
\end{array}\right]\left[\begin{array}{c}
\psi_{m-1,k}^{A1}\\
\psi_{m,k}^{B1}
\end{array}\right],\label{eq:matrix equality}
\end{align}
which is equivalent to Eq.~\eqref{tmatrix for leads}.

Similar steps can be taken to build the transfer matrix for the bilayer
region where $1\leq m\leq p$,

\begin{align*}
\left(\epsilon_{k}-V_{1}\right)\psi_{m,k}^{A1} & =-\eta\psi_{m,k}^{B1}-\xi\psi_{m+1,k}^{B1}-t_{\perp}\psi_{m,k}^{B2}\\
\left(\epsilon_{k}-V_{1}\right)\psi_{m,k}^{B1} & =-\eta^{\ast}\psi_{m,k}^{A1}-\xi^{\ast}\psi_{m-1,k}^{A1}\\
\left(\epsilon_{k}-V_{2}\right)\psi_{m,k}^{A2} & =-\eta\psi_{m,k}^{B2}-\xi\psi_{m+1,k}^{B2}\\
\left(\epsilon_{k}-V_{2}\right)\psi_{m,k}^{B2} & =-\eta^{\ast}\psi_{m,k}^{A2}-\xi^{\ast}\psi_{m-1,k}^{A2}-t_{\perp}\psi_{m,k}^{A1},
\end{align*}
from which we obtain Eq.~\eqref{eq:tmatrix for bilayer} in matrix
form.

By imposing the boundary conditions for setup $1\rightarrow1$, and
defining $\mathbb{M}=\left(\mathbb{T}_{BL}\right)^{p}$, we can re-write
Eq.~\eqref{M matrix} as
\begin{align}
\left[\begin{array}{c}
\psi_{p,k}^{A1}\\
\psi_{p+1,k}^{B1}
\end{array}\right] & =\mathcal{M}_{1\rightarrow1}\left[\begin{array}{c}
\psi_{0,k}^{A1}\\
\psi_{1,k}^{B1}
\end{array}\right]\,,
\end{align}
or equivalently,
\begin{align}
\Psi_{k}^{1}\left(p+1\right) & =\mathcal{M}_{1\rightarrow1}\Psi_{k}^{1}\left(1\right)\,,
\end{align}
where
\[
\mathcal{M}_{1\rightarrow1}=\frac{1}{\mathbb{M}_{44}}\left[\begin{array}{cc}
\mathbb{M}_{11}\mathbb{M}_{44}-\mathbb{M}_{14}\mathbb{M}_{41} & \mathbb{M}_{12}\mathbb{M}_{44}-\mathbb{M}_{14}\mathbb{M}_{42}\\
\mathbb{M}_{21}\mathbb{M}_{44}-\mathbb{M}_{24}\mathbb{M}_{41} & \mathbb{M}_{22}\mathbb{M}_{44}-\mathbb{M}_{24}\mathbb{M}_{42}
\end{array}\right]\,.
\]
Using the boundary condition for the setup $1\rightarrow2$, we obtain,
after similar steps,
\begin{align}
\Psi_{k}^{2}\left(p+1\right) & =\mathcal{M}_{1\rightarrow2}\Psi_{k}^{1}\left(1\right)\,,\label{eq:27}
\end{align}
where
\[
\mathcal{M}_{1\rightarrow2}=\frac{1}{\mathbb{M}_{24}}\left[\begin{array}{cc}
\mathbb{M}_{24}\mathbb{M}_{31}-\mathbb{M}_{34}\mathbb{M}_{21} & \mathbb{M}_{24}\mathbb{M}_{32}-\mathbb{M}_{34}\mathbb{M}_{22}\\
\mathbb{M}_{24}\mathbb{M}_{41}-\mathbb{M}_{44}\mathbb{M}_{21} & \mathbb{M}_{24}\mathbb{M}_{42}-\mathbb{M}_{44}\mathbb{M}_{22}
\end{array}\right]\,.
\]

The last step is to represent wave amplitudes $\Psi_{k}^{1\left(2\right)}\left(m\right)$
in the eigenbasis of the transfer matrix of the leads. The characteristic
equation for the eigenvalue problem $\mathbb{T}_{L}\zeta_{k}=\lambda\zeta_{k}$
reads, 

\begin{equation}
\left(\xi_{k}\eta_{k}^{\ast}\right)\lambda^{2}-\left(\varepsilon^{2}-|\xi_{k}|^{2}-|\eta_{k}|^{2}\right)\lambda+\xi_{k}^{\ast}\eta_{k}=0\,,\label{eq: characteristic of T}
\end{equation}
yielding two eigenvalues,

\begin{align}
\lambda_{\pm} & =\frac{1}{\xi_{k}\eta_{k}^{\ast}}\left(\varepsilon^{2}-|\xi_{k}|^{2}-|\eta_{k}|^{2}\pm\sqrt{\left(\varepsilon^{2}-\delta_{+}^{2}\right)\left(\varepsilon^{2}-\delta_{-}^{2}\right)}\right)\,,
\end{align}
where $\delta_{\pm}=|\xi_{k}|\pm|\eta_{k}|$, corresponding to the
normalized eigenvectors

\begin{equation}
\zeta_{k}^{\pm}=\frac{1}{\sqrt{2}}\left(\begin{array}{c}
1\\
\frac{-\varepsilon}{\xi_{k}^{\ast}+\eta_{k}^{\ast}\lambda_{\pm}}
\end{array}\right)\,.\label{eq: eigenvalues od T}
\end{equation}

A mode with positive (negative) group velocity is considered to be
the right-moving(+) (left-moving(-)) mode. Recalling that in the leads
$|\lambda|=1$ and using Bloch theorem for the Pristine graphene $\lambda=e^{iq\left(k,\varepsilon\right)}$,
where $q\left(k,\varepsilon\right)$ is the conjugate momentum in
$\mathbf{a}_{1}$(propagating) direction, and plugging the latter
expression into Eq. \eqref{eq: characteristic of T} we obtain the
mode group velocity in the propagating direction as:

\begin{equation}
v_{g}=\frac{d\varepsilon}{dq}=\frac{-1}{\varepsilon}Im\left(\xi_{k}\eta_{k}^{\ast}\lambda\left(k,\varepsilon\right)\right)\,.
\end{equation}

\bibliographystyle{apsrev4-1}
\bibliography{hadi-T_BLG}

\end{document}